# Developing a diagnostic e-module for testing conceptual understanding of Symmetry and Gauss's Law

Ashok Kumar Mittal[a)]

*Department of Physics, University of Allahabad, Allahabad, India 211002*

We discuss an approach for developing a diagnostic e-module, which can easily identify the precise nature of deficiencies in conceptual understanding of a topic, using Symmetry and Gauss's Law as an example. The topic is divided into suitably chosen concept categories. A large number of True/False questions are prepared for each concept category, which can be answered using the concepts in that category. The module presents seven randomly chosen questions from each of the concept categories. Only when each of the questions from a concept category is answered correctly, will it be inferred that the concept category has been understood properly. A diagnostic module based on this approach can reliably and efficiently identify the conceptual deficiencies. After taking remedial measures, the student can take the diagnostic test again till all conceptual deficiencies are removed.

## I. INTRODUCTION

Physics is a concept based subject. Experts regard with favor the attitude of making conceptual analysis of a problem before looking for solutions using equations and formulae[1]. Physics Education Research has shown that a large fraction of students at various levels of learning lack the correct conceptual understanding for almost all topics in Physics[2]. A substantial part of Physics Education Research has been concerned with this problem[3-17]. Various interactive-engagement instruction strategies have been developed to improve conceptual understanding. Several assessment tools for measuring conceptual understanding of different topics have also been developed. The effectiveness of the instruction strategies are evaluated using these conceptual understanding assessment tools before and after instruction. Some of the important conclusions from these studies are:

(i)     Traditional tests fail to reveal lack of conceptual understanding.

(ii)    Traditional lectures fail to make much improvement in conceptual understanding.

(iii)   Interactive-engagement instruction strategies yield considerable improvement in conceptual understanding.

However, the learning gains are still much below hundred percent showing that there are several students whose conceptual deficiencies remain, despite the best instructional efforts. In this paper, using the topic of Symmetry and Gauss's Law as an example, we present an approach to developing diagnostic e-modules for identifying the precise nature of conceptual deficiencies in a student. These modules can be used to ensure that each student develops a complete conceptual understanding about a topic.

The first step in creating the module is to divide the topic into well identified concept categories. Thereafter, for each concept category, a large class of True/False (T/F) questions is created. It should be possible to answer every question associated with a concept category by invoking the concepts from that category. Every question should be tagged to only one concept category.

This approach differs from prevailing conceptual tests in the following respects:

(i) Each question in our diagnostic module is associated with only one concept category.

(ii) The module consists of T/F type questions instead of Multiple Choice Questions (MCQ's).

(iii) The number of questions associated with each concept category in the module is very large (at least fifty for each concept category)

(iv) The diagnostic test presented to a student by the module consists of randomly chosen seven questions from each category.

(v) The questions are presented one at a time. The module provides immediate feedback, informing whether the answer is correct or wrong. This allows a student to think about the reasons for a wrong answer before proceeding further.

(vi) A student, who answers all the seven questions correctly, is diagnosed as having understood the concepts in that category.

(vii) A student, who is unable to answer all the questions correctly, is expected to take remedial measures and then take the test again. The module will present a fresh set of seven questions. The student must persist till successful.

(viii) The test may be taken at any time, any number of times, one category at a time or all the categories simultaneously.

In Sec II we discuss the concept categories into which we divide the topic of Symmetry and Gauss's Law, based on conceptual analysis of the type of questions a student is expected to be able to answer after instruction on the topic. With each concept category, we give an example of a question that can be answered by using the concepts in that category. The ability to answer all questions of this type for each identified concept category can be regarded as the conceptual learning goals for the topic. In Sec III we explain our approach for designing a diagnostic e-module which can reliably and efficiently ascertain whether a student has achieved this goal. In Sec IV we describe the advantages of this diagnostic module by comparing it with the conceptual test developed by Singh[5]. In Sec V we provide our conclusions.

## II. IDENTIFICATION OF CONCEPT CATEGORIES

In order to investigate student understanding of Symmetry and Gauss's Law, Singh[5] developed a set of 25 Multiple Choice Questions (MCQs). In this paper the concepts investigated were classified into nine categories. These were listed along with the question numbers which tested a particular category. It was also mentioned that the categories were not mutually exclusive; that a question could be listed against more than one category; that classification was based on students' difficulties and not on the basis of how an expert would categorize the concepts involved.

The notion of concept has been analyzed extensively in Physics Education Research literature[3]. Here we take a concept to mean a unit of mental thought in terms of which people organize their understanding of any topic. A concept category is a collection of concepts that are invoked to solve a class of problems.

Endless variety of questions can be asked around any topic. However, as a part of a curriculum, the concepts and type of problem solving abilities that a student is expected to learn about a topic are limited. In this limited context, it is possible to systematically list a set of concept categories and develop a large class of problems associated with each category which can be solved using the concepts from that category.

For the topic of Symmetry and Gauss's Law as a topic in a standard undergraduate course in Electricity and Magnetism, we identified the concept categories as follows: (1) Statement of Gauss's Law (2) Meaning of electric flux + Statement of Gauss's Law (3) Symmetry (4) Symmetry + Principle of Superposition (5) Additive property of Electric Flux + Statement of Gauss's Law + Symmetry (6) Determination of Electric Field using Guass's Law and Symmetry (7) Gauss's Law + Symmetry + explicit use of the Superposition Principle.

Gauss's Law states that the total outward electric flux over a closed surface is proportional to the total charge enclosed by the surface. The concept of electric flux and the principle of superposition are implicit in the Gauss's Law. However, a class of questions can be answered using the statement of the Gauss's Law without invoking explicitly the principle of superposition or the meaning of electric flux. An example question in this category is: "A and B are two intersecting closed surfaces. A charge $q_1$ is enclosed by A, but not by B. Charges $q_2$ and $q_3$ lie inside B, but not A. Both A and B enclose a charge $q_4$. A charge $q_5$ is outside both A and B. What is the total outward electric flux over A?" A student who does not understand the meaning of electric flux, but knows the statement of Gauss's Law can correctly answer that the electric flux over A will be $(q_1 + q_4)/\varepsilon_0$.

Another class of questions can be answered correctly by using the Statement of Gauss's Law along with explicitly invoking the meaning of electric flux. For example, consider the question: "Positive charge Q is distributed uniformly over the surface of a thin-walled cubic insulating (non-conducting) box. S is the surface of a sphere, which lies inside the box. Centre of S coincides with the centre of the box. Can the electric field point inward at every point on the surface of S?" As S is a closed surface which encloses no charge, from Gauss's Law it follows that the total outward electric flux over S is zero. From the meaning of electric flux it can be inferred that if the electric field points inward at every point of S, the total outward electric flux would be negative. The contradiction shows that the electric field cannot point inward at every point of S.

Use of symmetry arguments in conjunction with the Gauss's Law forms an important part of the instruction on Gauss's Law. For diagnostic purposes it is desirable to know whether a student can recognize symmetry and use it to draw inferences. Concept category (3) is identified with questions that can be answered by symmetry considerations alone – where, for the given problem there is no way of distinguishing between two or more directions or between two and more points or between two and more surfaces. An example question in this category is: "Charge is distributed uniformly over the surface of a thin-walled cubic insulating (non-conducting) box. What is the electric field at the centre of the cube?" The electric field at the centre of the cube has to be zero because for this problem there is no way of distinguishing between any direction and its opposite direction; if the electric field points in one direction there is no reason why it should not point in the opposite direction. This geometry based argument does not invoke the Coulomb's Law or the principle of superposition. The answer holds independent of the law relating electric field to charge distribution. The question can also be answered using a superposition based argument – the charge distribution is divided into pairs, each pair consisting of symmetrically placed charges for which the electric field at the centre of the cube cancels, and then using the principle of superposition to conclude that the field at this point is zero. Pepper *et al*[6] have emphasized the need for distinguishing between geometry based symmetry arguments and superposition based symmetry arguments. Most problems can be solved by both these methods. In order to test conceptual understanding of geometry based symmetry arguments, it is necessary to include in this category questions that cannot be answered using superposition based arguments. An example of such a question is: "S is an arbitrary scalar entity. V is a vector field related to S by some unknown law. S is distributed uniformly over a straight line AB. C is the midpoint of AB. CD is a line perpendicular to AB. What is the direction of V at any point P on CD?" The direction of V has to be along CD or DC, because if V is inclined at an angle $\theta$ with respect to CD there is no reason why it should not be inclined at an angle $-\theta$. This result is valid even if V is determined by a law for which the principle of superposition does not hold, for example, if V is determined by all of S concentrated at its centre.

The concept category (4) is identified with questions which cannot be answered using symmetry without invoking the principle of superposition also. An example of a question in this category is: "Two spheres of the same radius R, have their centers at the points (-d, 0, 0) and (d, 0, 0). Positive charge Q is distributed uniformly over the surface of each of the spheres. Another charge Q is placed at an arbitrary point P. What is the direction of the electric field at the origin O?" By symmetry it is inferred that the electric field at O due to the two charged spheres must be zero. Then using the principle of superposition it is concluded that the direction of the electric field at O must be along PO. It should be noted that the inverse square dependence of the electric force between two point charges is not needed. It is possible to infer from symmetry alone that the direction of the electric force between point charges must be along the line joining the point charges. Thus only symmetry and the principle of superposition are required for answering the question. To ensure that a student has understood this point, questions of the following type are included in this category: "S is an arbitrary scalar entity. V is a vector field related to S by some unknown law which satisfies the principle of superposition. Two spheres of the same radius R, have their centers at the points (-d, 0, 0) and (d, 0, 0). S is distributed uniformly over the surface of each of the spheres. S is also placed at an arbitrary point P. What is the direction of V at the origin O?" The contribution to V at the origin O from S on the two spheres is zero by symmetry. Again, by symmetry, the contribution to V at O due to S at P must be along PO. Since V satisfies the principle of superposition, direction of V at O must be along PO.

In the concept category (5) use is made of the additive property of flux (flux over a surface is equal to the sum of fluxes over its parts), in addition to the statement of Gauss's Law and Symmetry. An example question in this category is: "Two charges, Q, are placed at the points ( d, 0, 0) and (-d, 0, 0). If b > d, what is the electric flux over the hemisphere defined by $\{x^2 + y^2 + z^2 = b^2, x > 0\}$?" For the given charge distribution it is not possible to distinguish between this hemisphere and the hemisphere $\{x^2 + y^2 + z^2 = b^2, x < 0\}$. Therefore symmetry implies that the fluxes over these two hemispheres must be equal. By additive property, the flux over the sphere $\{x^2 + y^2 + z^2 = b^2\}$ will equal the sum of fluxes over the two hemispheres. However, by Gauss's Law the flux over the sphere is $2Q/\varepsilon_0$. Hence the flux over either hemisphere is $Q/\varepsilon_0$.

The concept category (6) deals with questions in which symmetry allows the flux over a closed surface to be expressed as the product of the electric field and some area. The electric field can then be determined easily using the Gauss's Law. An example question in this category is: "A triangular prism is placed, with its axis perpendicular to a uniformly charged infinite sheet having surface charge density σ. The mid-point of the axis lies in the sheet. B is a point on one of the triangular faces which is parallel to the sheet. Can the prism be used as a Gaussian surface to determine the electric field at B?" By symmetry of the charge distribution the electric field at every point must be perpendicular to the infinite sheet. Therefore the flux over each of the faces parallel to the axis of the prism is zero. As the two faces parallel to the infinite charged sheet are equidistant from the sheet, by symmetry the magnitude E of the electric field at every point on these two faces is the same. If A is the area of each of the two triangular faces of the prism, the total flux over the closed surface of the prism is 2EA. This must equal the total charge enclosed divided by $\varepsilon_0$. The total charge enclosed is σA. Hence E = $\sigma/(2\varepsilon_0)$ at every point on the two triangular faces. Thus the specified triangular prism can be used as a Gaussian surface for determining the electric field at the point B.

Questions in the concept category (7) require explicit use of the superposition principle along with Gauss's Law and Symmetry. An example question in this category is: "Charge Q is uniformly distributed over a solid sphere of radius R. A point charge Q is placed at a point P distance 2L from the centre of the sphere. M is the mid-point of the line joining P to the centre. If L > R, what is the electric field at M?" By principle of superposition the electric field at M is the sum of electric field due to the charged solid sphere and the electric field due to the point charge at P. As the point M is outside both the spheres, using symmetry and Gauss's Law it is possible to show that the electric field at M due to the charged solid sphere is the same as that due to a point charge Q placed at the centre of the sphere. The electric field at the mid-point between two equal point charges has to be zero by virtue of symmetry. Hence it follows that the electric field at M due to the charged solid sphere and the point charge is zero.

In addition to the above categories, we have also introduced a category: 'Preliminary concepts'. These include the concepts of charge, electric flux, closed surface, inside, outside and outward normal. The concept of flux further requires the concepts of electric field, normal to a surface, scalar product of vectors and surface integral. It is assumed that at the level of the module on Gauss's Law and Symmetry, all these preliminary concepts are well understood. This category is included to test whether this assumption is valid.

III. DESIGNING A DIAGNOSTIC MODULE

Several tests, based on Multiple Choice Questions (MCQ's) have been developed to test the conceptual understanding of different topics. These tests are widely used as pre and post test to compare the efficacy of different instruction strategies. They have to satisfy stringent statistical criterion for individual items of the test and for the overall test[7,8]. They are difficult and time-consuming to develop.

Though our diagnostic module also focuses on conceptual understanding, its purpose is not to evaluate and compare individual students, student populations or instruction strategies. The purpose is to diagnose whether an individual student has learnt a well identified set of concepts. So the questions in the diagnostic module do not have to be tested statistically for validity and reliability by administering it over a large population. The only requirement is that the concepts required to be invoked for answering a particular question are properly identified so that the question is tagged to a unique concept category.

Sophisticated tools have been developed for proper design of MCQ based tests and for drawing inferences from the data generated by the answers[9,10]. MCQ's are useful for comparing the overall performance of individual students or of student populations. However, they are not satisfactory for a diagnostic tool. Quite often a question does not admit sufficiently purposive distracters. The requirement of three or more distracters usually leads to combining more than one concept. To make useful inference about conceptual understanding, each MCQ has to be analyzed individually. Randomly chosen options can lead to totally wrong inferences about the conceptual understanding. For example, consider the following question[5]:

Choose all of the following physical variables that are vectors: (i) Electric field (ii) Electric Flux (iii) Charge

(a) (i) only

(b) (i) and (ii) only

(c) (i) and (iii) only

(d) (ii) and (iii) only

(e) (i), (ii) and (iii)

In this MCQ three independent concepts – electric field, electric flux and electric charge – have been combined into one question, for no other reason than the requirement of a fixed number of options. Because of this requirement, the options are also not exhaustive. What if a student thinks that electric field is a scalar, electric flux is a vector and charge is a scalar? Not finding this option a student may randomly choose the option (c). From this choice the inference about the student's concepts will be totally erroneous.

Instead it is much better to ask three T/F type questions:

1. True or False: Electric Field is a vector.

2. True or False: Electric Flux is a vector.

3. True or False: Charge is a vector.

The only problem with a T/F type question is that the probability of answering a question correctly without proper understanding is higher than for MCQ. This problem can be overcome by asking a large number of T/F questions for each concept category. The probability of randomly answering seven questions is less than 1%. So a diagnostic module should present at least seven questions for each identified concept category in order to reliably ascertain that the concept category has been learnt. The number of questions available for a concept category must be much larger than seven, so that even after taking the diagnostic test many times a student does not get to memorize the answers. In our diagnostic module there are about 650 questions with at least 50 questions in each concept category.

The diagnostic module is very flexible and can easily be adapted to different instruction goals and instructors' perceptions about the conceptual misunderstandings amongst their students.

IV. ADVANTAGES OF THE DIAGNOSTIC MODULE

We discuss the advantages of the diagnostic module by comparing it to the conceptual test consisting of 25 MCQ's designed by Singh[5] to explore student understanding of symmetry and Gauss's Law. Problem 13 of the conceptual test is:

The surface of a thin-walled cubic insulating (nonconducting) box is given a uniformly distributed positive surface charge. Which of the following can be inferred about the electric field everywhere inside the insulating box due to this surface charge using Gauss's law?

(a) Its magnitude everywhere inside must be zero.

(b) Its magnitude everywhere inside must be nonzero but uniform (the same)

(c) Its direction everywhere inside must be radially outward from the center of the box.

(d) Its direction everywhere inside must be perpendicular to one of the sides

(e) None of the above

According to Singh[5] this question addresses the following four concepts: (i) Recognizing the symmetry of the charge distribution (ii) Difference between the electric field and the electric flux (iii) Recognizing the symmetry to determine if it is easy to exploit Gauss's law or exploiting Gauss's law to determine the electric field (iv) Electric field inside hollow nonconducting objects with different charge distribution. The link between the question and these concepts must have been established on the basis of interviews with students. However from the option chosen by a student, it is not possible to draw any reliable inference about the conceptual understanding of the student. Many students tend to answer questions on the basis of superficial resemblances to other problems they may have come across, rather than by careful reasoning. Option (a) may have been chosen due to superficial resemblance of the problem with that of the conducting box or with that of a uniformly charged insulating spherical shell. It is difficult to think of any reasons for options (b) and (d) except as random choice. Option (c) would be chosen due to the common misconception that a charge distribution can be replaced by a charge at its centre. This misconception is an overgeneralization from the observation that extended bodies are often treated as point objects. Many students fail to appreciate the conditions under which this can be done and the necessary arguments. From the option chosen, one may make some guess about the misconceptions of a student, but it is not possible to attribute the conceptual deficiency to a particular category listed against this question. Due to the possibility of random choice, the correct option (e) also does not allow the inference that the student has understood the concepts involved. Also, the student may have encountered the problem earlier and remembered the result without fully understanding the concepts.

In our categorization, this question belongs to the category (6): 'Determination of electric field using Gauss's Law and Symmetry'. We conjecture that any person who can answer each of the seven randomly chosen questions from this category will also be able to answer this question correctly. If a student is successful with the first five categories but fails in the sixth, it can be inferred reliably that the student understands the Gauss's Law and Symmetry concepts well, but still is unable to put them together to tackle this particular concept category.

Clearly, the diagnostic module can efficiently and reliably identify the precise nature of deficiency in conceptual understanding of a student. After taking remedial measures a student can take the diagnostic test again with another set of randomly chosen questions. Some people understand a concept quickly from few examples. Others need more examples. As our diagnostic module contains a large number of examples, it is reasonable to expect that any motivated student can achieve hundred per cent gain in conceptual understanding in a short time. As soon as a student manages to answer seven randomly chosen questions belonging to each of the categories, no further effort needs to be spent towards conceptual understanding of the topic. Thus the diagnostic module is a very efficient tool for directing a student's efforts in an optimal manner for acquiring correct conceptual understanding.

Although the diagnostic module does not directly offer any instruction, it should be regarded as a useful tool that encourages active learning and interactive teaching[11-17]. It satisfies many goals that have been considered desirable by researchers in Physics Education. It encourages a student centric approach to education, allowing a student to move at own pace and seek help as best suited to own needs. It focuses a student's attention on one concept at a time. Immediate feedback makes a student aware about inadequacy of understanding. It forces a student to confront misconceptions and gives an opportunity to modify the conceptions so as to bring them in line with the correct answer. The module gives repeated exposure to the same concepts in a variety of contexts till a large number of all correct responses indicate that no conceptual misunderstanding remains. The conceptual understanding so acquired after a mental struggle and reinforced in several different contexts is likely to remain firmly embedded in the long term memory. The module helps students see the relationships between different concepts and how to combine them in different contexts. This helps them build appropriate knowledge hierarchies.

The diagnostic module can be used as a supplement to any instruction strategy from the traditional to the highly interactive. It does not require highly motivated and competent instructors. It can be used by a student in the most indifferent instruction environment.

## V. CONCLUSION

We have presented an approach for development of a reliable and efficient diagnostic module for identifying deficiencies in conceptual understanding of Symmetry and Gauss's Law. The topic is divided into well defined concept categories. A large number of T/F questions are developed such that each question is identified with exactly one concept category. The diagnostic test consists of seven randomly chosen questions from each of the concept categories. The probability, of answering each of these questions from a concept category correctly without understanding, is less than one per cent. Failure to answer all the seven questions from a category implies that there is a conceptual deficiency in that category and the student is required to take focused remedial measures and thereafter to take the diagnostic test again till each of the seven randomly chosen questions from each of the categories is correctly answered. The diagnostic module can optimally direct the efforts of a student to efficiently achieve full conceptual understanding of a topic.

[a]Electronic Mail: mittals79@hotmail.com